\documentclass[%
reprint,
amsmath,amssymb,
aps,
prb,
floatfix,
]{revtex4-2}

\usepackage{graphicx}% Include figure files
\usepackage{dcolumn}% Align table columns on decimal point
\usepackage{subfigure}
\usepackage{bm}% bold math
\usepackage{url}
\usepackage[T2A,T1]{fontenc}
\usepackage[utf8]{inputenc}
\usepackage{float}
\usepackage{amsmath,amssymb}
\usepackage{lmodern}
\usepackage{mathtools}
\usepackage{bm}
\usepackage{physics}
\usepackage{braket}
\usepackage{xcolor}
\usepackage{graphicx}
\usepackage{calc}
\usepackage{hyperref}
\hypersetup{
	colorlinks=true,
	linkcolor=blue,
	citecolor=blue,
	filecolor=magenta,      
	urlcolor=blue,
}

\begin{document}
	
	\title{Analytic T-Matrix Formalism for Helmholtz Resonators}
	
	\author{Egor~V.~Domoratskii$^{1,2}$}
	\email{These authors contributed equally to this work}
	
	\author{Polina~A.~Medvedeva$^{2}$}
	\email{These authors contributed equally to this work}
	
	\author{Dmitrii~N.~Maksimov$^{3,4}$}
	
	\author{Lana~A.~Glazko$^{2}$}
	
	\author{Alexey~A.~Proskurin$^{2}$}
	
	\author{Nikolay~S.~Solodovchenko$^{1,2,5}$}
	
	\author{Mingzhao~Song$^{1}$}
	
	\author{Yong~Li$^{2,6}$}
	\email{yongli@tongji.edu.cn}
	
	\author{Mihail~I.~Petrov$^{2}$}
	\email{m.petrov@metalab.ifmo.ru}
	
	\author{Andrey~A.~Bogdanov$^{1,2}$}
	\email{a.bogdanov@hrbeu.edu.cn}
	
	\affiliation{$^{1}$ Qingdao Innovation and Development Center, Harbin Engineering University, Qingdao 266000, China}
	
	\affiliation{$^{2}$ School of Physics and Engineering, ITMO University, St. Petersburg 197101, Russia}

	\affiliation{$^{3}$ Kirensky Institute of Physics, Federal Research Center KSC SB RAS, Krasnoyarsk, Russia}
	
	\affiliation{$^{4}$ IRC SQC, Siberian Federal University, Krasnoyarsk 660041, Russia}
	
	\affiliation{$^{5}$ Ioffe Institute, St. Petersburg 194021, Russia}
	
	\affiliation{$^{6}$ Institute of Acoustics, School of Physics Science and Engineering, Tongji University, Shanghai 200092, China}

	\begin{abstract}
		We develop an analytic $T$-matrix formalism for two-dimensional circular acoustic Helmholtz resonators with thin rigid walls. The central idea is to use an aperture projection operator to obtain a compact expression for the $T$-matrix with explicit coefficients describing coupling and decoupling between external fields and cavity resonances. An equivalent formulation providing much faster convergence uses an aperture-to-Fourier map and an edge-adapted basis for the aperture flux. The method extends directly to multiple apertures of arbitrary angular widths and positions. The calculated scattering spectra agree with full-wave numerical simulations. For single- and double-aperture resonators, we decompose the $T$-matrix into even and odd parts under inversion to identify the channels associated with Willis coupling. We show how internal losses enable direction-dependent scattering and absorption when inversion symmetry is broken, while reciprocity ensures equal extinction for opposite incidence. Finally, we show how shifting the expansion origin redistributes radiation among multipole channels, enabling more compact $T$-matrix representations.
	\end{abstract}
	
	\maketitle

	\section{Introduction}
	
	Helmholtz resonators are fundamental, extensively studied acoustic systems~\cite{rayleigh1945theoryofsound,morse1968theoretical,kinsler2000fundamentals,ingard1953theoryanddesign,pierce1989acoustics}. Their simple geometry supports strongly confined resonances at wavelengths much larger than their dimensions, making them building blocks for acoustic devices and canonical models of resonant scattering, wave confinement, and radiation losses. This simplicity also facilitates development and validation of analytic scattering models.
	
	The study of Helmholtz resonators has a long history, dating back to the pioneering work of Helmholtz~\cite{helmholtz1863lehre} and Rayleigh's classical theory~\cite{rayleigh1916theory}, which related the resonance frequency to the geometry of the cavity and its opening. Subsequent theoretical and experimental studies have explored the effects of cavity shape and volume, aperture dimensions, and neck length and profile. Howe~\cite{howe1976helmholtz} further developed resonator theory, while Monkewitz and Nguyen-Vo~\cite{monkewitz1985response} established a matched-asymptotic treatment of externally excited resonators. Selamet and Lee~\cite{selamet2003helmholtz} studied neck extensions, and Mercier \textit{et al.}~\cite{mercier2017influence} derived analytical relations between neck shape and resonance frequencies. More recently, Brandão and Schnitzer~\cite{brandao2020asymptotic} developed a systematic asymptotic theory incorporating cavity geometry, radiation damping, and thermoviscous effects.
	
	Modern interest is driven by acoustic metamaterials and metasurfaces, which widely employ Helmholtz resonators as subwavelength building blocks~\cite{deymier2013acoustic,cummer2016controlling,ma2016acoustic,assouar2018acoustic,huang2023soundabsorbing,yang2017sound,gao2022acoustic,zhang2020engineering,romerogarcia2020design}. An early demonstration by Fang \textit{et al.}~\cite{fang2006ultrasonic} used a one-dimensional resonator array to realise an ultrasonic metamaterial with a negative effective bulk modulus. Applications now include low-frequency sound absorption, noise suppression, wavefront manipulation, and unconventional effective acoustic properties~\cite{cummer2016controlling,ma2016acoustic,assouar2018acoustic,cox2017acoustic,yang2017sound}. Critical coupling enables perfect and broadband absorption in deeply subwavelength resonant structures~\cite{romerogarcia2020design,jimenez2017rainbow}. Coupled resonators and periodic arrays also enable collective resonances~\cite{Smith2022_part_1, Smith2022_part_2, krasikova2024acoustic,krasikova2023metahouse}, tailored acoustic band structures~\cite{krasikova2024broadband,piva2026designing,elmalki2026defective}, and directional scattering~\cite{hesford2010acoustic,elford2011matryoshka,krynkin2011scattering,montiel2020planar} or absorption~\cite{krasikova2023metahouse}. All these developments require fast, robust computational methods and advanced analytical frameworks that efficiently capture higher-order resonances, multipolar scattering, and coupling between radiation channels beyond the conventional low-frequency monopole approximation.
	
	Despite this geometric simplicity, analytical modelling remains challenging because the aperture introduces mixed boundary conditions, edge singularities, and coupling between cavity modes and exterior radiation. Opening a closed, lossless cavity transforms its discrete real eigenfrequencies into complex resonances, with frequency shifts and radiative linewidths that depend nontrivially on aperture geometry~\cite{gadylshin1996scattering,gadylshin1997existence,gadylshin2003scattering}. Eliminating the interior and exterior fields reduces the problem to an aperture integral equation involving frequency-dependent, non-local boundary operators, related to the Dirichlet-to-Neumann and Neumann-to-Dirichlet maps~\cite{nedelec2001acoustic,colton2019inverse}. Approaches to this problem include aperture integral equations~\cite{norris1991dynamic,yang1991shear,norris1993elastic}, eigenfunction~\cite{montiel2020planar,moran2016embedding} and multipole~\cite{ustimenko2025optimal,ustimenko2026lattice} expansions, and asymptotic methods~\cite{brandao2020asymptotic,gadylshin1997existence,burrows1985waves,mohring2000helmholtz,vanbladel1969coupling,smith2022asymptotics}. In particular, Smith \textit{et al.}~\cite{smith2022asymptotics} combined multipole expansions and matched asymptotics to obtain low-frequency scattering coefficients for two-dimensional Helmholtz resonators with different wall thicknesses, describing their resonances and complex poles. The corresponding scalar Neumann problem also appears in the electromagnetic analysis of a slotted conducting cylinder~\cite{nosich1977electromagnetic}.
	
	\begin{figure}[t]
		\includegraphics[width=\linewidth]{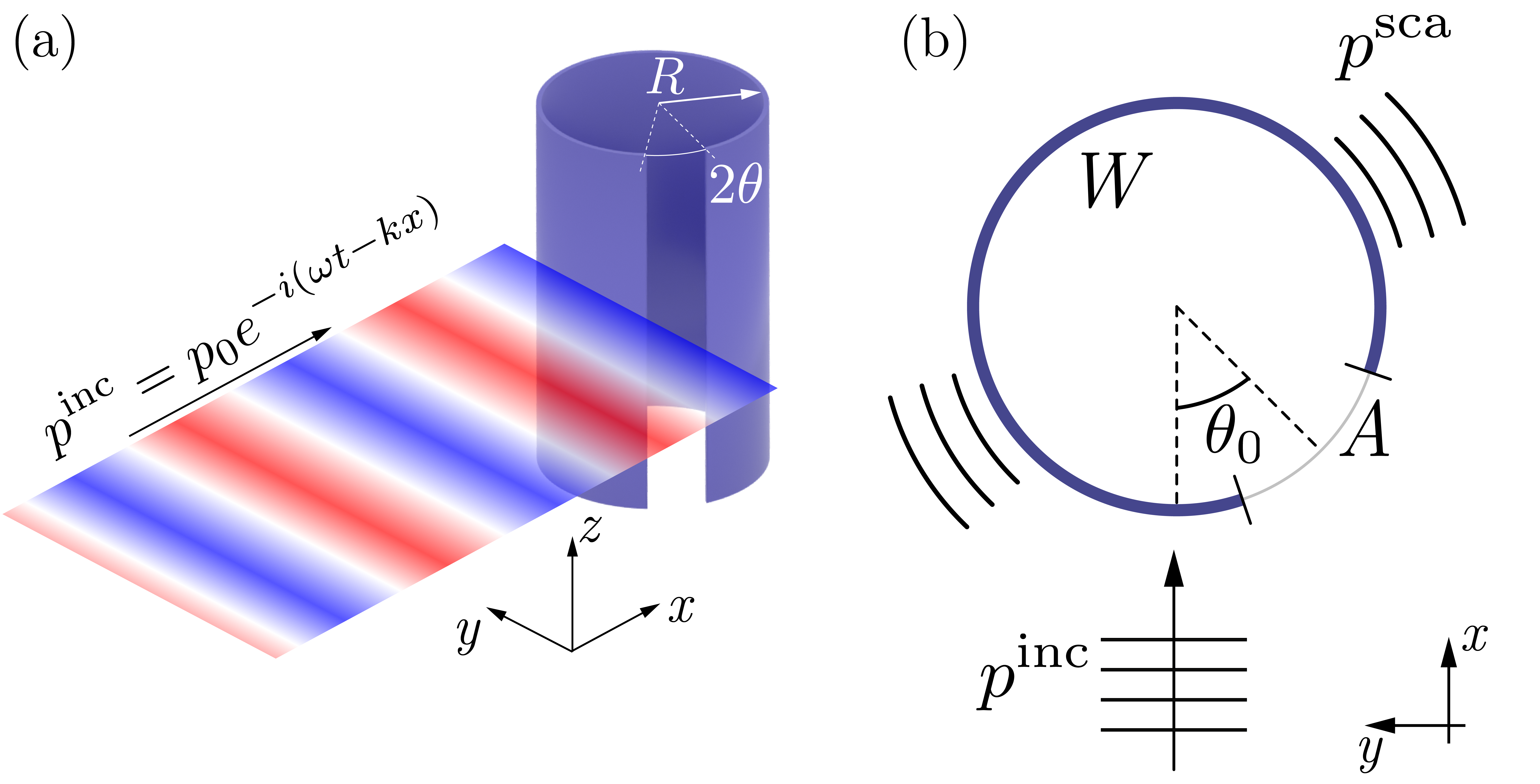}
		\caption{
			Geometry of the open cylindrical Helmholtz resonator and the scattering configuration.
			(a) An infinitely long, hard-wall cylindrical shell of radius $R$, containing a longitudinal slit, is illuminated by an incident plane pressure wave $p^\mathrm{inc}$.
			(b) Cross-sectional view of the resonator. The slit has a full angular aperture width of $2\theta$, while the remaining cylindrical boundary is sound-hard. The incident wave propagates along the $x$ direction toward the aperture.
		}
		\label{fig:figure_1_scattering_geometry}
	\end{figure}
	
	The transition-matrix ($T$-matrix) formalism is a powerful tool for describing scattering by individual resonators and their assemblies. It relates incident and scattered cylindrical- or spherical-wave amplitudes, allowing arbitrary excitations to be treated without repeatedly solving the boundary-value problem. Translation and addition theorems extend this description to finite clusters and periodic arrays. Waterman first established the electromagnetic framework~\cite{waterman1965matrix}, extended it to acoustics~\cite{waterman1969newformulation}, and reviewed subsequent developments~\cite{waterman2009tmatrix}. Matrix formulations of multiple scattering were developed by Peterson and Str\"om~\cite{peterson1974matrix} and systematically discussed by Martin~\cite{martin2011multiple}. Tools such as \textit{acoustotreams}~\cite{ustimenko2026acoustotreams} facilitate calculations for resonator ensembles, while analytic continuation of the $T$-matrix connects scattering spectra to complex resonance poles. Existing descriptions include finite-frequency aperture and modal formulations~\cite{norris1993elastic,moran2016embedding,nosich1977electromagnetic,wilks2024generalized}, related Green-function treatments of Dirichlet scattering~\cite{rao1999plane}, effective boundary models for perforated cylinders~\cite{skvortsov2025acoustic}, and asymptotic multipole and $T$-matrix expressions~\cite{smith2022asymptotics,piva2026designing}. Despite these advances, establishing an explicit operator-level $T$-matrix representation that connects aperture geometry, complex resonance spectra, and multipole coupling for arbitrary aperture configurations beyond low-frequency or narrow-aperture approximations remains underexplored.
	
	Here, we develop an analytic $T$-matrix formalism for two-dimensional circular acoustic Helmholtz resonators with thin rigid walls and multiple apertures of arbitrary angular widths and positions. An aperture projection operator separates the closed-cylinder response from aperture-induced scattering and yields explicit coefficients for coupling incident waves to cavity resonances and decoupling the cavity modes into outgoing scattering channels. The same aperture equation determines the complex resonance poles. An equivalent Galerkin formulation uses an aperture-to-Fourier map and an edge-adapted basis for the aperture flux. This formulation converges substantially faster than a direct Fourier expansion, particularly for narrow apertures, and the calculated scattering spectra agree closely with full-wave finite-element simulations.
	
	The rest of the paper is organized as follows. In Sec.~\ref{sec:theoretical_model}, we derive the $T$-matrix. In Sec.~\ref{sec:single_Helmholtz_resonator}, we analyze the scattering spectra and complex resonance poles as functions of aperture width. In Sec.~\ref{sec:multipole_analysis}, we examine inversion symmetry, Willis coupling, and direction-dependent scattering and absorption in single- and double-aperture resonators. Finally, in Sec.~\ref{Sec:sct_origin}, we show that shifting the expansion origin yields more compact multipole representations of the scattering response.

	\section{T-Matrix of Helmholtz Resonator}
	\label{sec:theoretical_model}
	
	\subsection{Problem Formulation}
	
	We consider scattering of a plane wave by two-dimensional circular acoustic Helmholtz resonators with thin rigid walls (Fig.~\ref{fig:figure_1_scattering_geometry}). The aperture of angular width $2\theta$, centred at $\theta_0$, occupies the angular interval $A = [-\theta + \theta_0, \theta_0 + \theta]$. The pressure fields satisfy the scalar Helmholtz equation
	\begin{equation}
		\nabla^2 p + k^2 p = 0,
	\end{equation}
	with $k = \omega/c$, where $\omega$ is the angular frequency and $c$ is the speed of sound, assumed identical inside and outside the resonator. Here and in what follows, we assume the harmonic time dependence as $\exp(-i \omega t)$. The incident and scattered pressure fields $p^{\mathrm{inc}}$ and $p^{\mathrm{sca}}$ can be expanded into a series of cylindrical harmonics:
	\begin{equation}
		\begin{aligned}
			p^{\mathrm{inc}}(r,\varphi;\omega) = \sum_{m=-\infty}^{+\infty}
			a_m J_m(kr) e^{i m \varphi},
			\\
			p^{\mathrm{sca}}(r,\varphi;\omega) = \sum_{m=-\infty}^{+\infty}
			b_m H_m(kr) e^{i m \varphi},
		\end{aligned}
		\label{eq:incident_and_scattered_fields_expansions}
	\end{equation}
	where $J_m(kr)$ and $H_m(kr) \equiv H_m^{(1)}(kr)$ are the $m$-th order Bessel and Hankel functions of the first kind, respectively. The total exterior pressure field is $p^\mathrm{ext} = p^\mathrm{inc} + p^\mathrm{sca}$. The interior pressure field must remain finite at $r = 0$ and can therefore be expanded solely in regular cylindrical harmonics:
	\begin{equation}
		p^{\mathrm{int}}(r,\varphi;\omega) = \sum_{m=-\infty}^{+\infty}
		c_m J_m(kr) e^{i m \varphi}.
		\label{eq:internal_fields_expansion}
	\end{equation}
	The coefficients $a_m$ in Eq.~\eqref{eq:incident_and_scattered_fields_expansions} specify the incident field, whereas the unknown coefficients $b_m$ in Eq.~\eqref{eq:incident_and_scattered_fields_expansions} and $c_m$ in Eq.~\eqref{eq:internal_fields_expansion} describe the scattered and interior fields, respectively.
	
	By definition, the T-matrix reads:
	\begin{equation}
		\mathbf{b} = \hat{T}(\omega) \mathbf{a},
		\label{eq:connection_between_amplitudes}
	\end{equation}
	where the vectors $\mathbf{b}=\{b_m\}$ and $\mathbf{a}=\{a_m\}$ are composed of the scattered and incident complex amplitudes. For the closed cylinder with hard walls, the T-matrix ($\hat T_0$) is diagonal as the angular momentum of the incident sound is preserved:
	\begin{equation}
		(T_0)_{mm'}=-\frac{J'_m(kR)}{H'_m(kR)}\delta_{mm'}.    
	\end{equation}
	
	The cylindrical boundary $r=R$ is divided into the hard-wall region $W$ and the aperture region $A$, as shown in Fig.~\ref{fig:figure_1_scattering_geometry}(b). On the rigid wall $W$, the normal derivatives of the interior and exterior pressure fields vanish, whereas
	across the aperture $A$, both the pressure and its normal derivative are continuous. The boundary conditions therefore read
	\begin{equation}
		\left\{
		\begin{aligned}
			\left.\partial_r p^{\mathrm{int}}\right|_W
			&= \left.\partial_r p^{\mathrm{ext}}\right|_W = 0,\\
			\left.p^{\mathrm{int}}\right|_A
			&= \left.p^{\mathrm{ext}}\right|_A,\\
			\left.\partial_r p^{\mathrm{int}}\right|_A
			&= \left.\partial_r p^{\mathrm{ext}}\right|_A.
		\end{aligned}
		\right.
		\label{eq:slit_boundary_conditions}
	\end{equation}

	\subsection{Projection-Operator Formalism}
	
	To impose the boundary conditions separately on the aperture and the rigid wall, we introduce the aperture projection operator $\hat P_A$ through its action on an angular boundary function $g(\varphi)$:
	\begin{equation}
		\hat{P}_A[ g(\varphi)]=
		\begin{cases}
			\displaystyle
			g(\varphi), \ \ \varphi \in A \\
			\displaystyle
			0, \ \ \varphi \in W \\
		\end{cases}.
		\label{eq:aperture_projection_operator}
	\end{equation}
	It can be also written as $\hat{P}_A[g(\varphi)] = \Theta(\theta - |\varphi - \theta_0|) g(\varphi)$, where $\Theta$ is the Heaviside step function. Thus, $\hat{P}_A$ retains the function on the aperture $A$ and sets it to zero on the rigid wall. The projection operator is Hermitian $\hat{P}_A^\dagger = \hat{P}_A$. In the Fourier representation, it can be written as 
	\begin{equation}
		(P_A)_{mm'}
		= e^{-i(m-m')\theta_0}
		\begin{cases}
			\displaystyle \frac{\theta}{\pi},
			& m=m', \\[4pt]
			\displaystyle
			\frac{\sin[(m-m')\theta]}{\pi(m-m')},
			& m\ne m'.
		\end{cases}
		\label{eq:aperture_projection_operator_matrix_element}
	\end{equation}
	
	We introduce the normalized radial derivative on the cylindrical boundary,
	\begin{equation}
		q(\varphi)
		= \frac{1}{k}\partial_r p^{\mathrm{int}}(R,\varphi)
		= \frac{1}{k}\partial_r p^{\mathrm{ext}}(R,\varphi),
		\label{eq:q_derivation}
	\end{equation}
	which vanishes on the rigid wall and can be nonzero only on the aperture. Its Fourier expansion and coefficient vector are
	\begin{equation}
		q(\varphi)=\sum_m q_m e^{im\varphi},
		\qquad
		\mathbf{q}=\{q_m\}.
	\end{equation}
	The aperture-support condition is therefore $\mathbf{q}=\hat{P}_A\mathbf{q}$.

	The boundary conditions~\eqref{eq:slit_boundary_conditions}, together with Eqs.~\eqref{eq:incident_and_scattered_fields_expansions}, \eqref{eq:internal_fields_expansion}, and~\eqref{eq:q_derivation}, yield the following relations between the complex amplitude vectors $\mathbf{a}$, $\mathbf{b}$, $\mathbf{c}$, and $\mathbf{q}$:
	\begin{equation}
		\begin{aligned}
			\hat{J}'\mathbf{c} &= \mathbf{q},\\
			\hat{J}'\mathbf{a}+\hat{H}'\mathbf{b}
			&= \mathbf{q},\\
			\hat{P}_A
			\left(
			\hat{J}\mathbf{a}+\hat{H}\mathbf{b}
			-\hat{J}\mathbf{c}
			\right) &= 0.
		\end{aligned}
		\label{eq:boundary_conditions_operator}
	\end{equation}
	Here we introduced the diagonal operators
	\begin{equation}
		\begin{aligned}
			\hat{J} &= \mathrm{diag}[J_m(kR)],
			& \hat{H} &= \mathrm{diag}[H_m(kR)],\\
			\hat{J}' &= \mathrm{diag}[J_m'(kR)],
			& \hat{H}' &= \mathrm{diag}[H_m'(kR)].
		\end{aligned}
		\label{eq:radial_operators}
	\end{equation}
	Eliminating $\mathbf{c}$ and $\mathbf{b}$, applying the Bessel-Hankel Wronskian identity, and using $\mathbf{q}=\hat{P}_A\mathbf{q}$ yields the closed operator expression for the $T$-matrix:
	\begin{equation}
		\hat{T}
		= \hat{T}_0
		+\underbrace{(\hat{H}')^{-1}\hat{P}_A}_{\hat{W}_{\mathrm{out}}}
		\underbrace{\left[\hat{P}_A(\hat{H}'\hat{J}')^{-1}\hat{P}_A\right]^{-1}}_{\hat{K}^{-1}}
		\underbrace{\hat{P}_A(\hat{H}')^{-1}}_{\hat{W}_{\mathrm{in}}}.
		\label{eq:t_matrix}
	\end{equation}
	
	The first term describes scattering by the closed hard-wall cylinder, while the second accounts for the aperture-induced response associated with cavity resonances. This factorized structure is analogous to {\it temporal coupled-mode theory}, which separates direct background scattering from a resonant contribution involving coupling to resonant modes and their decoupling into outgoing channels~\cite{fan2003temporal,ruan2012temporal,zhou2016perfect,maksimov2025temporal}. Here, $\hat{T}_0$ provides the direct background, while $\hat{W}_{\mathrm{in}}$ and $\hat{W}_{\mathrm{out}}$ describe coupling of incident waves into the aperture-mediated response and decoupling into outgoing radiation channels, respectively. The poles of $\hat{K}^{-1}$ determine the open-cavity resonances. Details of the derivation are given in Sec.~I of the Supplemental Material~\cite{supplemental_material}. Equation~\eqref{eq:t_matrix} applies to arbitrary aperture widths, including large openings, and continuously recovers both limiting cases. As $\theta\to0$, the aperture closes and $\hat{T}\to\hat{T}_0$, recovering the closed hard-wall cylinder. As $\theta\to\pi$, the wall disappears, $\hat{P}_A\to\hat{I}$, and the two terms cancel, yielding $\hat{T}\to0$.
	
\begin{figure}[t]
		\includegraphics[width=\linewidth]{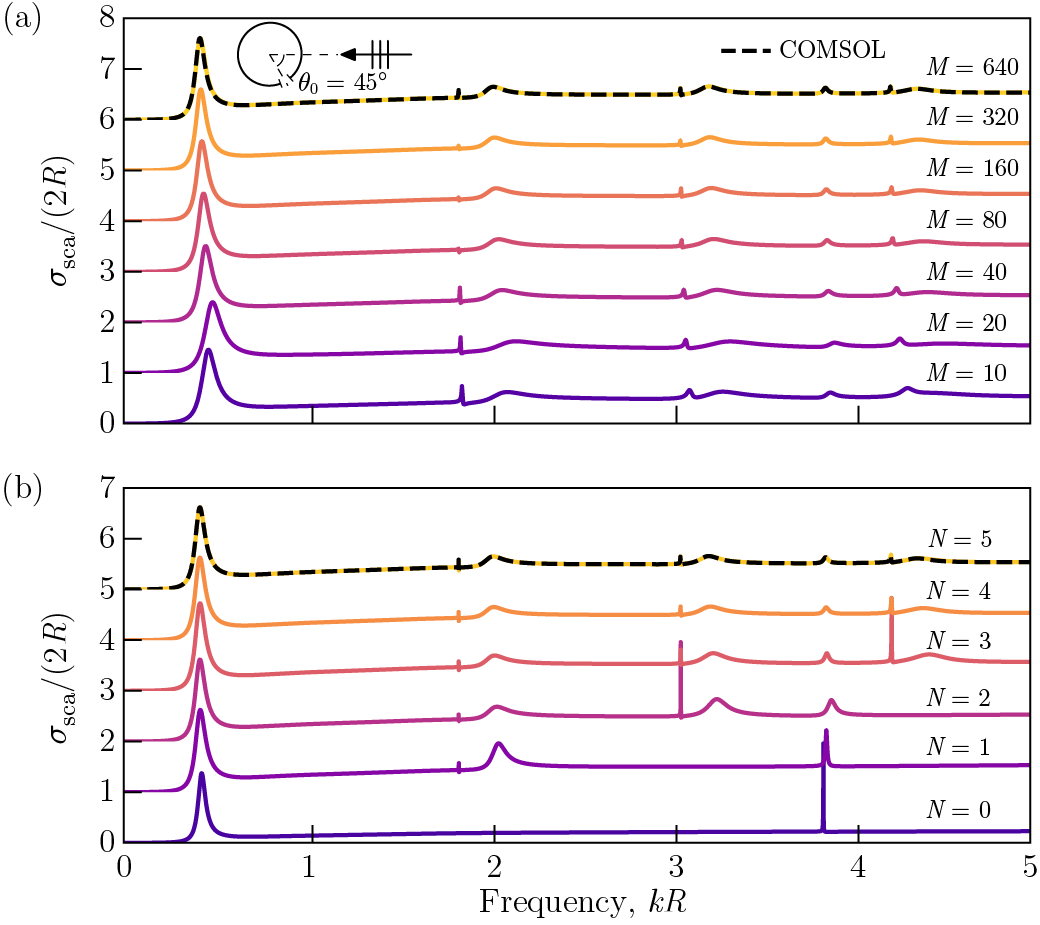}
		\caption{
			Convergence of the normalized scattering cross section $\sigma_{\mathrm{sca}}/(2R)$ for $\theta_0=45^\circ$.
			(a) Projection-operator representation with Fourier cutoff $M$.
			(b) Edge-adapted aperture-basis representation with aperture cutoff $N$ and $L = 2 $ slit functions. Solid curves show the analytical results, and dashed black curves show the finite-element COMSOL reference. Spectra are vertically offset for clarity.
		}
		\label{fig:figure_2_convergence}
	\end{figure}

	For numerical evaluation, we retain a finite number of angular harmonics by truncating the expansions to $|m|\le M$. The incident and scattered amplitude vectors then contain $2M+1$ components, and the $T$-matrix has dimensions $(2M+1) \times (2M+1)$. The same cutoff also limits the angular sums in the aperture operator and the Fourier expansion of $q(\varphi)$. Figure~\ref{fig:figure_2_convergence}(a) shows the convergence of the normalized scattering cross section $\sigma_{\mathrm{sca}} / (2R)$ as $M$ increases. For a plane wave with pressure amplitude $p_0$ incident at angle $\theta_0$ with respect to the aperture position [Fig.~\ref{fig:figure_1_scattering_geometry}(a)], the cross section is
	\begin{equation}
		\sigma_{\mathrm{sca}}
		= \frac{4}{k|p_0|^2}\,
		\mathbf{a}^{\dagger}
		\hat{T}^{\dagger}\hat{T}
		\mathbf{a},
		\label{eq:scattering_cross_section}
	\end{equation}
	where $a_n=p_0 i^n\exp(-in\theta_0)$ are the incident-wave coefficients. The spectra converge slowly toward the finite-element reference obtained using COMSOL Multiphysics, with substantial discrepancies persisting even for $M$ about 100--300.

	The slow convergence arises from several factors. (i) First, the internal angular harmonics contributing to the scattering response need not be limited to the few external channels that dominate the radiated field. In the present implementation, however, the same cutoff $M$ controls the internal and external angular harmonics. (ii) Second, the aperture flux is localized within the opening and exhibits integrable inverse-square-root singularities at its edges. Representing this distribution using smooth global Fourier harmonics requires progressively larger cutoffs as the aperture narrows. (iii) Finally, although the operator $\hat{P}_A$ is a projector, for the truncated basis $\hat{P}_A^2\neq\hat{P}_A$. A finite Fourier expansion cannot represent a field that vanishes exactly outside the aperture. Consequently, truncation does not preserve the aperture-support condition and can distort the coupling and decoupling coefficients represented by $\hat{W}_{\mathrm{in}}$ and $\hat{W}_{\mathrm{out}}$. The aperture-space inverse in Eq.~\eqref{eq:t_matrix} therefore requires a numerical treatment that accounts for this constraint. More details on convergence are provided in Secs.~II and~III of the Supplemental Material~\cite{supplemental_material}.

	\subsection{Aperture-Basis Representation}
	
	To account explicitly for the edge singularities and reduce the number of harmonics required to represent the aperture flux $q(\varphi)$, we introduce an auxiliary coordinate $u\in[0,\pi]$ that parametrizes the aperture:
	\begin{equation}
		\varphi(u) = \theta_0 + \theta \cos u.
		\label{eq:aperture_parametrization}
	\end{equation}
	The endpoints $u=0$ and $u=\pi$ correspond to the aperture edges. We define the transformed flux $\widetilde{q}(u)=q(\varphi)|d\varphi/du|$. With this parametrization, the edge singularities are incorporated into the transformation, leaving the behaviour of $\widetilde{q}(u)$ regular. We expand this amplitude in an independent aperture basis:
	\begin{equation}
		\widetilde{q}(u)
		= \sum_{\ell=0}^{L}d_\ell\psi_\ell(u),
		\qquad
		\psi_\ell(u)
		=
		\begin{cases}
			1, & \ell=0,\\
			\sqrt{2}\cos(\ell u), & \ell\ge1.
		\end{cases}
		\label{eq:aperture_basis}
	\end{equation}
	These functions satisfy
	$\pi^{-1}\int_0^\pi
	\psi_\ell(u)\psi_{\ell'}(u)\,du=\delta_{\ell\ell'}$.
	Thus, $L+1$ aperture harmonics describe the regular part of the flux.
	
	This choice of aperture basis follows established approaches to aperture integral equations and edge-weighted Chebyshev expansions in resonator and circular-arc scattering~\cite{lee1971waveinduced,norris1993elastic,moran2016embedding}. Related Galerkin methods have also been applied to multilayer split-ring resonators~\cite{montiel2020planar}. A closely related edge-weighted Chebyshev-Galerkin formulation was developed for electromagnetic scattering by a thin-walled slotted cylinder with TE polarization~\cite{scharstein2006near}. The present representation treats arbitrary aperture widths without the narrow-aperture assumptions used in asymptotic methods~\cite{smith2022asymptotics}. 
	
	The Fourier coefficients of the flux are related to the aperture coefficients $\mathbf{d}=\{d_\ell\}$ by
	\begin{equation}
		\mathbf{q}=\hat{D}\mathbf{d},
		\qquad
		D_{m\ell}
		= \frac{1}{2\pi}\int_0^\pi
		\psi_\ell(u)e^{-im\varphi(u)}\,du.
		\label{eq:aperture_flux_mapping}
	\end{equation}
	The operator $\hat{D}$ maps aperture coefficients to the angular Fourier coefficients introduced in the preceding subsection. Since the flux vanishes outside the aperture, the exact mapping satisfies $\hat{P}_A\hat{D}=\hat{D}$.
	
	Substituting the aperture expansion in Eq.~\eqref{eq:aperture_flux_mapping} into the boundary conditions~\eqref{eq:boundary_conditions_operator} and applying Galerkin testing with the same aperture basis gives
	\begin{equation}
		\hat{T}
		= \hat{T}_0
		+(\hat{H}')^{-1}\hat{D}
		\underbrace{
			\left[
			\hat{D}^{\dagger}
			(\hat{H}'\hat{J}')^{-1}\hat{D}
			\right]^{-1}
		}_{\hat{K}_D^{-1}}
		\hat{D}^{\dagger}(\hat{H}')^{-1}.
		\label{eq:t_matrix_aperture_basis}
	\end{equation}
	This expression retains the factorized structure of Eq.~\eqref{eq:t_matrix}. The operator $\hat{K}_D$ describes the aperture response, while $\hat{D}^{\dagger}(\hat{H}')^{-1}$ and $(\hat{H}')^{-1}\hat{D}$ describe coupling from incident channels and decoupling into outgoing channels, respectively.
	
	For external orders $|m|\le M$ and aperture orders $0\le\ell\le L$, the coupling matrix $\hat{D}$ has dimensions $(2M+1)\times(L+1)$. It does not have to be squared and can not be inverted. The retained $T$-matrix has dimensions $(2M+1)\times(2M+1)$, while only the aperture system with $L+1$ unknowns is solved.
	
	Another independent cutoff $N$ is introduced when evaluating the matrix element of $\hat K_D$:
	\begin{equation}
		(K_D)_{\ell\ell'}
		\simeq
		\sum_{n=-N}^{N}
		\frac{D_{\ell n}^{\dagger}D_{n\ell'}}
		{H_n'(kR)J_n'(kR)}.
		\label{eq:aperture_matrix_fourier_sum}
	\end{equation}
	Here, $n$ labels the angular harmonics of the combined interior and exterior response. The matrix $\hat{D}$ used in this sum has dimensions $(2N+1)\times(L+1)$, whereas $\hat{K}_D$ remains an $(L+1)\times(L+1)$ matrix. Increasing $N$ therefore facilitates calculating the aperture response without enlarging either the aperture system or the retained $T$-matrix.
	
	The sum in Eq.~\eqref{eq:aperture_matrix_fourier_sum} converges slowly with $N$. Its convergence can be accelerated by separating the leading large-order contribution and summing it analytically over all angular orders~\cite{scharstein2006near,moran2016embedding,abramowitz1965handbook,gradshteyn2014table}. Only the remaining correction is then truncated at $|n|\le N$. Details of the derivation and numerical implementation are provided in Sec.~III of the Supplemental Material~\cite{supplemental_material}.
	
	Figure~\ref{fig:figure_2_convergence}(b) shows the resulting convergence with $N$ for $M=10$ and a fixed aperture basis with $L = 2$. The spectra rapidly approach the finite-element reference. Already at $N=5$, the agreement is better than that obtained with the projection-operator formulation at $M\leq600$ in panel~(a). Convergence in $L$ must also be checked independently, since wider apertures and higher frequencies can require additional aperture functions.

	\section{Complex Poles and Scattering Spectra}
	\label{sec:single_Helmholtz_resonator}
	
	\subsection{Scattering Spectra}
	
	\begin{figure}[ht!]
		\includegraphics[width = \linewidth]{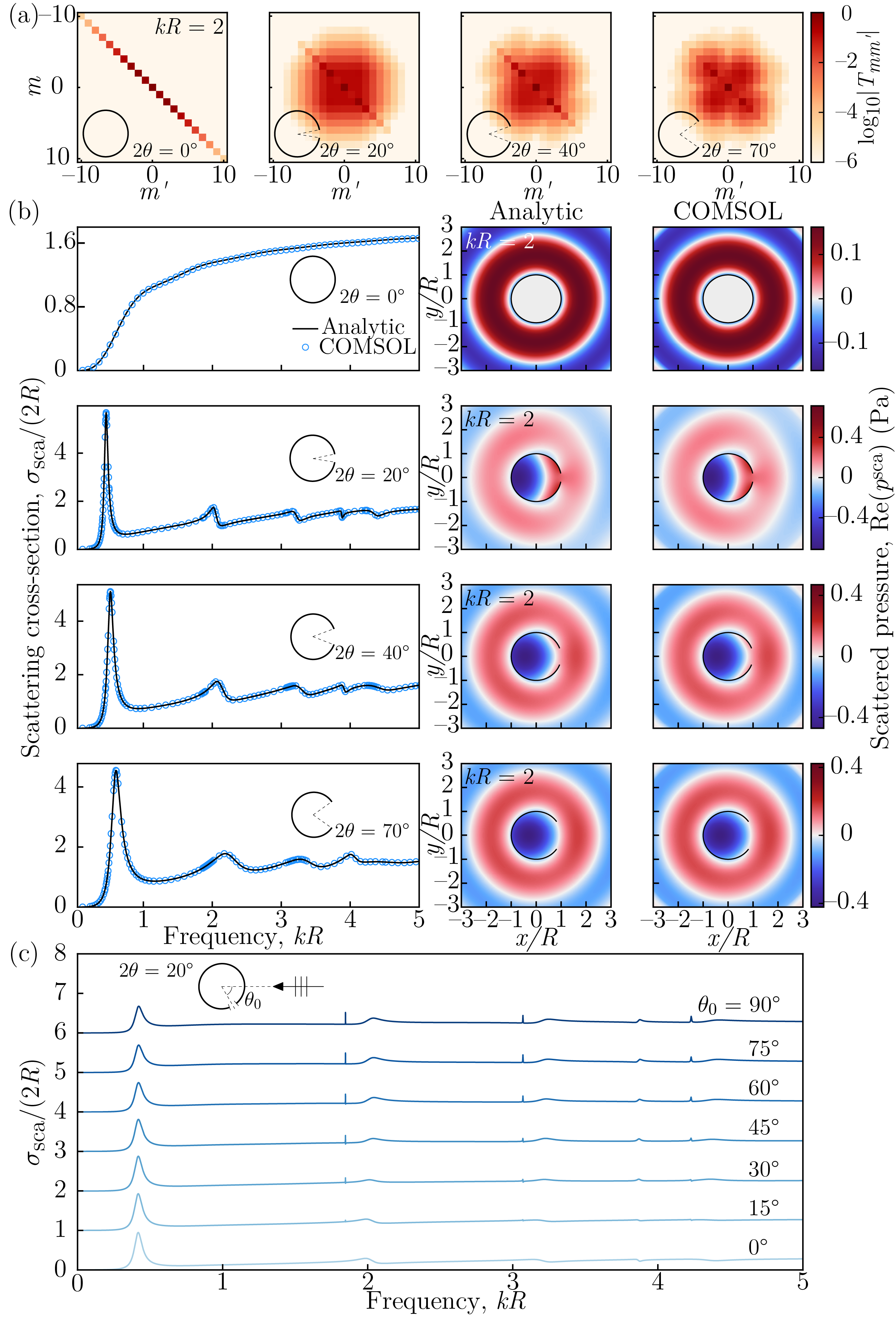}
		\caption{
			(a) Logarithmic $T$-matrix structure $\log_{10}|T_{mn}|$ at normalized frequency $\mathrm{Re}(kR) = 2$.
			(b) Normalized scattering cross-section $\sigma_\mathrm{sca}/(2R)$ as a function of the dimensionless frequency $kR$ for the four specific configurations in panel (a). Solid curves show the analytical results from Eq.~\eqref{eq:scattering_cross_section}, and open circles show the COMSOL Multiphysics results. The right-hand columns compare analytical and numerical maps of $\mathrm{Re}(p^\mathrm{sca})$ at $kR = 2$. Incidence angle is $\theta_0 = 0^\circ$.
			(c)  Dependence of the analytical spectra on the incidence anlge $\theta_0$ for $2\theta = 20^\circ$. The curves are vertically offset for clarity and illustrate changes in scattering spectra as $\theta_0$ increases.
		}
		\label{fig:figure_3_t_matrix_structure_and_scattering_spectra}
	\end{figure}
	
	We first examine how the aperture changes the scattering response of the circular resonator. Figure~\ref{fig:figure_3_t_matrix_structure_and_scattering_spectra} compares the closed cylinder with resonators having aperture widths $2\theta=20^\circ$, $40^\circ$, and $70^\circ$.  Figure~\ref{fig:figure_3_t_matrix_structure_and_scattering_spectra}(a) shows $\log_{10}|T_{mm'}|$ at $kR=2$. For the closed cylinder, the matrix is diagonal because angular momentum is conserved. Opening the slit breaks rotational symmetry and introduces off-diagonal elements, describing conversion between incident and outgoing angular orders. Their distribution changes with aperture width, showing how the opening modifies coupling between scattering channels. Figure~\ref{fig:figure_3_t_matrix_structure_and_scattering_spectra}(b) shows the corresponding scattering spectra. The closed cylinder has a relatively smooth response. Opening the aperture couples closed-cavity modes to the external radiation continuum, producing a low-frequency Helmholtz resonance and additional higher-frequency features. As the aperture widens, the Helmholtz resonance shifts to higher frequencies and broadens, consistent with established theory~\cite{bigg1982twodimensional,smith2022asymptotics}. The analytical results agree closely with finite element simulations, both in the scattering spectra and in the scattered pressure distributions at $kR=2$ shown in the two right-hand columns.
	
	Figure~\ref{fig:figure_3_t_matrix_structure_and_scattering_spectra}(c) shows the angular dependence of scattering for $2\theta=20^\circ$. The single-slit resonator has mirror symmetry about the slit axis, so its modes are symmetric or antisymmetric under reflection. Incidence along this axis, $\theta_0=0$ or $\pi$, excites only symmetric modes. For oblique incidence, additional narrow resonances corresponding to antisymmetric modes become visible. Their higher quality factors reflect weaker radiative coupling through the aperture, associated with smaller coupling overlap integrals. The incidence angle changes the excitation strengths of the resonances, while their complex frequencies remain unchanged. We next examine these complex frequencies to relate the spectral features to the resonance positions and radiative linewidths.

	\begin{figure}[t]
		\includegraphics[width = \linewidth]{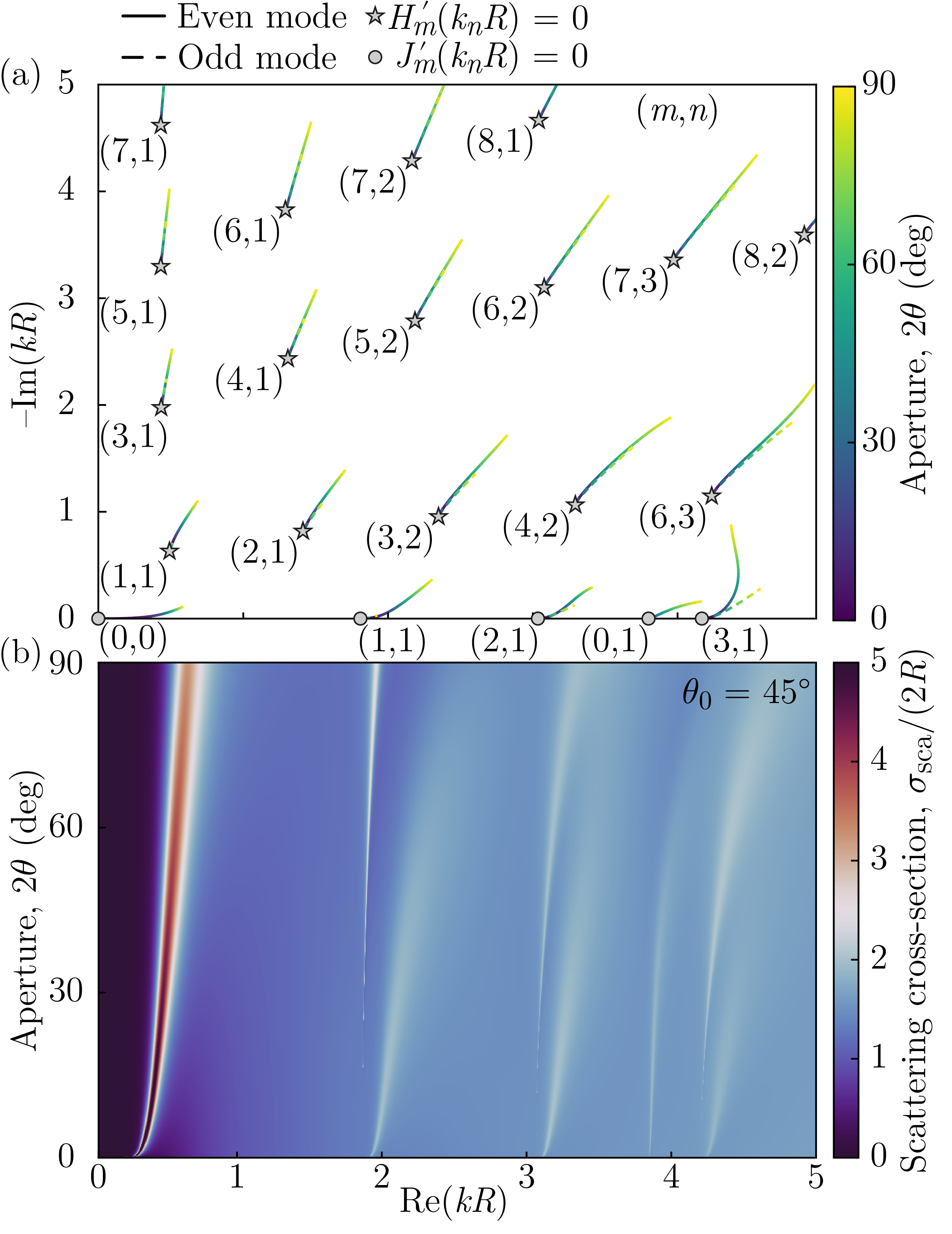}
		\caption{
			(a) Evolution of the complex resonance poles $k_nR$ with increasing aperture width $2\theta$, indicated by colour. Solid and dashed curves denote symmetric and antisymmetric modes, respectively. Grey circles and stars mark the internal and external modes of the closed resonator, satisfying $J_m'(kR)=0$ and $H_m'(kR)=0$. The labels $(m,n)$ denote the angular index and root number, with $(0,0)$ identifying the constant internal mode.
			(b) Normalized scattering cross section $\sigma_{\mathrm{sca}}/(2R)$ versus $kR$ and $2\theta$ for $\theta_0 = 45^\circ$.
		}
		\label{fig:figure_4_scattering_and_eigenfrequencies_maps}
	\end{figure}

	\subsection{Complex Poles}
	
	Figure~\ref{fig:figure_4_scattering_and_eigenfrequencies_maps}(a) shows the evolution of the complex poles $k_nR$ of the T-matrix as the aperture width $2\theta$ increases. The real parts determine the resonant frequencies, while the negative imaginary parts are eadiative decay rates. The corresponding linewidths are $\Gamma_n=-2\mathrm{Im}(\omega_n)$, where $\omega_n=ck_n$. The modes of the closed resonator are divided into internal modes satisfying $J_m'(kR)=0$ and external modes satisfying $H_m'(kR)=0$, marked by grey circles and stars, respectively. The labels $(m,n)$ denote the angular index and root number. The constant internal mode $(0,0)$ at $kR=0$ gives rise to the Helmholtz resonance when the aperture opens.
	
	For the closed resonator, cylindrical symmetry makes the modes with angular orders $m$ and $-m$ degenerate for $m\ne0$. The slit lifts this degeneracy, producing symmetric and antisymmetric modes, as discussed in the preceding subsection. Their evolution is shown by solid and dashed curves, respectively. Details of the numerical procedure are provided in Sec.~IV of the Supplemental Material~\cite{supplemental_material}. Figure~\ref{fig:figure_4_scattering_and_eigenfrequencies_maps}(b) shows how the scattering spectra evolve with aperture width. The spectral features approximately follow the real parts of the poles, extending the trends in Fig.~\ref{fig:figure_3_t_matrix_structure_and_scattering_spectra}(b). Narrow spectral features correspond to antisymmetric modes, while broader features are associated with symmetric modes.
	
	The fundamental Helmholtz pole approaches the origin as the aperture closes. Its leading narrow-aperture behaviour is
	\begin{equation}
		\begin{aligned}
			\mathrm{Re}(k_0R)
			&\sim [2\ln(2/\theta)]^{-1/2},\\
			\mathrm{Im}(k_0R)
			&\sim -\frac{\pi}{8}
			[2\ln(2/\theta)]^{-3/2}.
		\end{aligned}
		\label{eq:asymptotics}
	\end{equation}
	These expressions follow from the small-aperture and low-frequency expansion of the aperture operator using standard cylindrical-function asymptotics~\cite{morse1953methods}. The derivation is given in Sec.~V of the Supplemental Material~\cite{supplemental_material}. The frequency dependence recovers the classical result for two-dimensional resonators with small openings~\cite{burrows1985waves,norris1993elastic,bigg1982twodimensional}, also obtained by matched asymptotics~\cite{smith2022asymptotics}. As the aperture narrows, the radiative width decreases faster than the resonance frequency. The radiation quality factor therefore increases as
	\begin{equation}
		Q
		= \frac{\mathrm{Re}(k_0R)}
		{2|\mathrm{Im}(k_0R)|}
		\sim \frac{8}{\pi}\ln(2/\theta).
		\label{eq:helmholtz_quality_factor_asymptotics}
	\end{equation}
	This behaviour reflects the singular perturbation introduced by opening a closed cavity~\cite{gadylshin1996scattering,gadylshin1997existence,gadylshin2003scattering}. The increasing quality factor is closely related to the classical {\it harbor paradox}, when narrowing a harbor entrance can increase resonant amplification despite weakening its coupling to incident waves~\cite{miles1961harbor,mehaute1962closure}. The apparent divergence of $Q$ is resolved by accounting for viscous and thermal losses, which limit the quality factor and resonant amplification~\cite{ingard1953theoryanddesign,brandao2020asymptotic}.

	\begin{figure*}[t]
		\includegraphics[width=\linewidth]{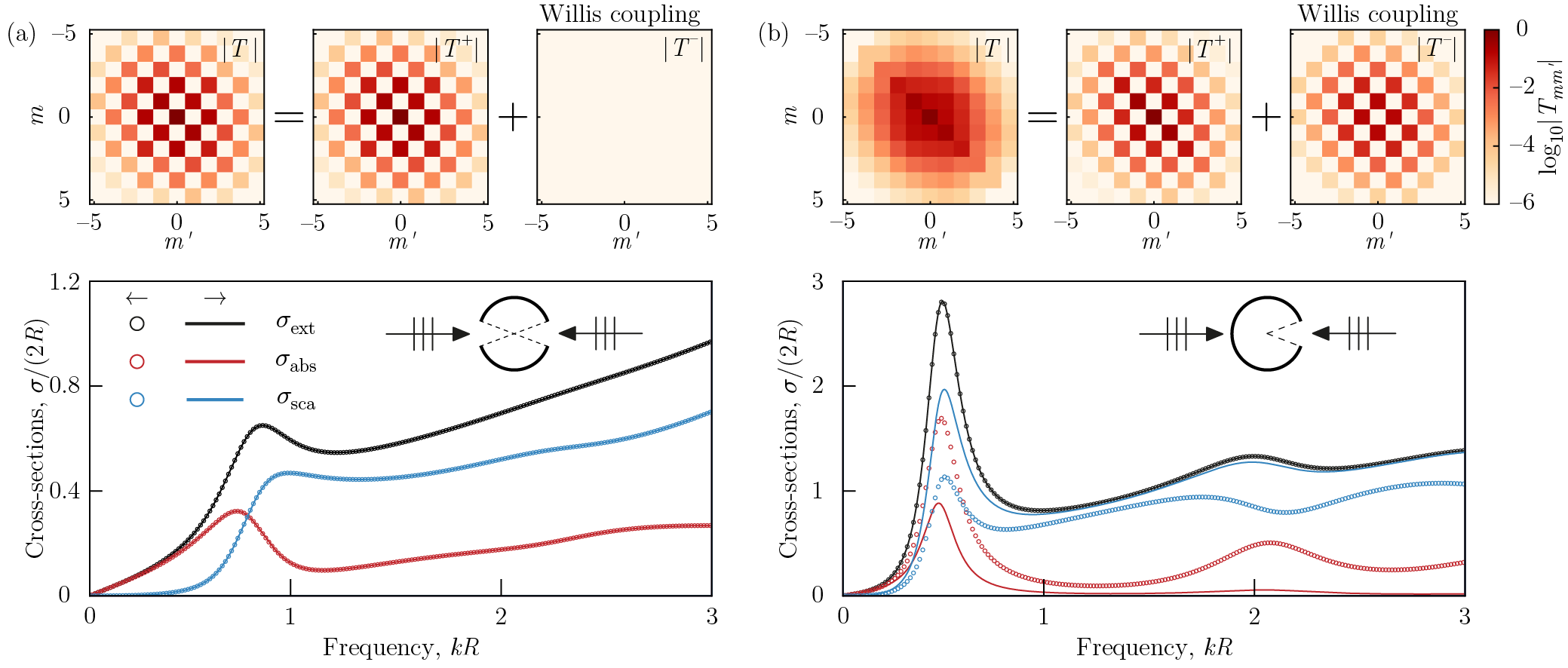}
		\caption{
			Inversion symmetry and directional response of
			resonators with (a) two identical, diametrically
			opposed slits and (b) a single slit.
			Each slit has width $2\theta=40^\circ$.
			Upper panels: $\log_{10}|T_{mm'}|$ and the
			corresponding inversion-symmetric and
			inversion-antisymmetric parts at $kR=1$.
			Lower panels: normalized extinction (black),
			absorption (red), and scattering (blue) cross
			sections. Solid lines and open circles denote
			rightward and leftward propagation, respectively.
			Internal losses are described by
			$k^{\mathrm{int}}=k(1+i\tan\delta)$ with
			$\tan\delta=0.1$.
		}
		\label{fig:figure_5_resonator_with_losses_and_two_slits}
	\end{figure*}

	\section{Inversion Symmetry and Willis Coupling}
	\label{sec:multipole_analysis}
	
	For a subwavelength acoustic scatterer without inversion symmetry, incident pressure can induce a dipole moment, while incident velocity can induce a monopole moment. These cross-responses constitute Willis coupling, the acoustic analogue of bianisotropy~\cite{sieck2017origins,quan2018maximum,su2018retrieval,melnikov2019acoustic}. Willis coupling in Helmholtz resonators was also discussed in Ref.~\cite{smith2022asymptotics}. Unlike nonreciprocal Willis coupling produced by a rotational flow bias~\cite{quan2021oddwillis}, the directional response considered below preserves reciprocity.
	
	We decompose the $T$-matrix into inversion-symmetric and inversion-antisymmetric parts:
	\begin{equation}
		\hat{T}^{\pm}
		= \frac{1}{2}
		\left(\hat{T}\pm\hat{\mathcal{P}}\hat{T}\hat{\mathcal{P}}\right),
		\qquad
		\mathcal{P}_{mm'}=(-1)^m\delta_{mm'}.
		\label{eq:expansion_of_t_matrix_into_parts}
	\end{equation}
	The inversion operator maps $\varphi$ to $\varphi+\pi$ and satisfies $\hat{\mathcal{P}}^2=\hat{I}$. Thus, $\hat{T}^{+}$ couples angular orders of the same inversion parity, while $\hat{T}^{-}$ couples opposite parities. Their nonzero elements occupy complementary checkerboard patterns.
	
	In the monopole-dipole approximation, the elements $T^{-}_{0,\pm1}$ and $T^{-}_{\pm1,0}$ describe Willis coupling after conversion to the pressure--velocity basis and appropriate normalization~\cite{su2018retrieval}. The same decomposition extends to higher multipoles: $T^{-}_{mm'}$ can be nonzero only when $m$ and $m'$ have opposite parity. These elements describe higher-order Willis-type scattering couplings, such as dipole--quadrupole and monopole-octupole coupling. If the geometry, material parameters, and loss distribution preserve inversion symmetry, then $\hat{\mathcal{P}}\hat{T}\hat{\mathcal{P}}=\hat{T}$ and $\hat{T}^{-}=0$, provided the expansion origin is the symmetry centre. Inversion symmetry therefore suppresses all opposite-parity couplings but allows coupling between different orders of the same parity.
	
	To examine dissipation, we use a uniform internal wavenumber $k^{\mathrm{int}}=k(1+i\tan\delta)$, with $\tan\delta=0.1$, while the exterior wavenumber $k=\omega/c$ remains real. The densities remain equal and real, and the walls remain rigid and lossless. Pressure and normal particle velocity are continuous across the apertures. This reciprocal model describes attenuation in the cavity fluid without explicitly treating viscous and thermal boundary layers.
	
	The scattering cross section is given by Eq.~\eqref{eq:scattering_cross_section}. In the same notation, extinction and absorption are
	\begin{equation}
		\begin{aligned}
			\sigma_{\mathrm{ext}}
			&= -\frac{4}{k|p_0|^2}
			\mathrm{Re}\!\left(
			\mathbf{a}^{\dagger}\hat{T}\mathbf{a}
			\right),\\
			\sigma_{\mathrm{abs}}
			&= \sigma_{\mathrm{ext}}-\sigma_{\mathrm{sca}}.
		\end{aligned}
		\label{eq:extinction_and_absorption}
	\end{equation}
	Equal extinction for opposite incidence follows from reciprocity~\cite{wang2023strong}.  Indeed, for opposite plane-wave directions the incident coefficients satisfy $a_{\leftarrow,m}=(-1)^m a_{\rightarrow,m}$ and $a_{\rightarrow,-m}=a_{\rightarrow,m}^{*}$. Together with the reciprocity condition $T_{mn}=(-1)^{m+n}T_{-n,-m}$, these relations give
	\begin{equation}
		\mathbf{a}_{\leftarrow}^{\dagger}
		\hat{T}\mathbf{a}_{\leftarrow}
		=
		\mathbf{a}_{\rightarrow}^{\dagger}
		\hat{T}\mathbf{a}_{\rightarrow},
	\end{equation}
	and hence equal extinction for opposite incidence. See the Supplemental Material for derivation details~\cite{supplemental_material}.

	The scattering cross section instead depends on the squared scattered amplitudes. Using Eq.~\eqref{eq:scattering_cross_section} and the parity decomposition gives
	\begin{equation}
		\begin{aligned}
			\Delta\sigma_{\mathrm{sca}}
			&\equiv
			\sigma_{\mathrm{sca}}^{\rightarrow}
			-\sigma_{\mathrm{sca}}^{\leftarrow}\\
			&= \frac{16}{k|p_0|^2}
			\mathrm{Re}\!\left[
			\mathbf{a}_{\leftarrow}^{\dagger}
			(\hat{T}^{+})^{\dagger}
			\hat{T}^{-}\mathbf{a}_{\rightarrow}
			\right].
		\end{aligned}
		\label{eq:directional_scattering_contrast}
	\end{equation}
	Directional contrast therefore arises from interference between the two parity contributions. A nonzero $\hat{T}^{-}$ permits this contrast but does not guarantee it. Since extinction is equal for both incidences,
	\begin{equation}
		\Delta\sigma_{\mathrm{ext}}=0,
		\qquad
		\Delta\sigma_{\mathrm{abs}}
		=-\Delta\sigma_{\mathrm{sca}}.
		\label{eq:cross_section_constraints}
	\end{equation}
	Internal losses thus allow different scattering and absorption for opposite incidence while preserving their sum. Without absorption, the integrated scattering cross sections must also coincide, although the angular scattering patterns can differ.

	Figure~\ref{fig:figure_5_resonator_with_losses_and_two_slits} compares two identical, diametrically opposed slits with a single slit. The two-slit resonator preserves inversion symmetry, so $\hat{T}^{-}=0$ and each cross section is identical for opposite incidence. The single slit breaks inversion symmetry and produces nonzero opposite-parity elements. Near its Helmholtz resonance at $kR\approx0.5$, incidence from the slit side gives stronger absorption and weaker scattering, while extinction remains unchanged, as required by Eq.~\eqref{eq:cross_section_constraints}.

	\section{Compact Multipole Representations through Origin Selection}
	\label{Sec:sct_origin}
	
	Shifting the expansion origin leaves the total scattering cross section unchanged but redistributes the scattered power among multipole channels. For a fixed incident plane wave, Eq.~\eqref{eq:scattering_cross_section} gives
	\begin{equation}
		\sigma_{\mathrm{sca}}=\sum_{m=0}^{\infty}\sigma_m.
		\label{eq:partial_scattering_cross_section}
	\end{equation}
	Here $\hat{T}$ is the full resonator matrix. The monopole contribution $\sigma_0$ contains only the angular harmonic $m=0$, while each higher-order cross-section $\sigma_m$ includes both $\pm m$ harmonics. Thus, $\sigma_1$ and $\sigma_2$ are the dipole and quadrupole contributions, respectively.
	
	We shift the origin horizontally (along the $x$ axis) from the cylinder centre by distance $d$ while keeping the resonator and incident wave fixed. For this particular displacement, Graf's addition theorem~\cite{abramowitz1965handbook} gives the translation matrix
	\begin{equation}
		U_{mn}(d)=J_{n-m}(kd).
		\label{eq:translation_operator_matrix_element}
	\end{equation}
	Both the incident and scattered amplitudes transform with $\hat{U}(d)$. For real $k$, the translation operator is unitary in the complete multipole basis, so the translated $T$-matrix is $\hat{T}'=\hat{U}(d)\hat{T}\hat{U}^{\dagger}(d)$ and the total scattered power is unchanged. In a  truncated basis, shifting the scattering origin is generally not unitary. Depending on $kR$ and $d/R$, some scattered power can be redistributed into truncated angular harmonics with $|m|>M$, causing the calculated total cross section to depend on the scattering origin offset.

	\begin{figure}[t]
		\includegraphics[width=\linewidth]{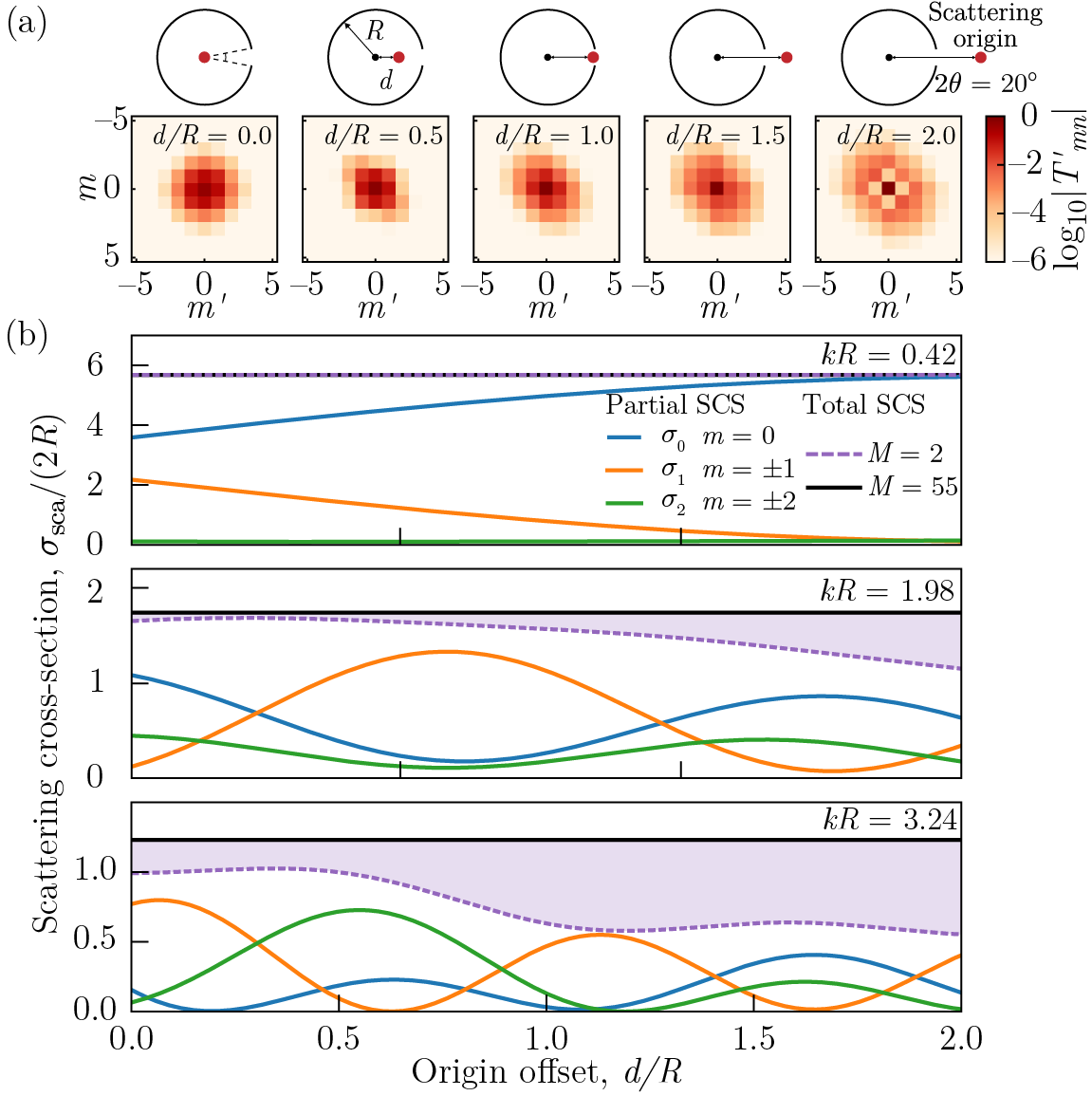}
		\caption{
			Effect of the expansion origin on scattering by a resonator with a single slit of width $2\theta=20^\circ$.
			(a) Maps of $\log_{10}|T'_{mn}|$ as the origin (red dot) moves along the slit axis at $kR = 0.42$.
			(b) Scattering cross sections normalized by $2R$ at $kR=0.42$, $1.98$, and $3.24$.
			Blue, orange, and green curves show $\sigma_0$, $\sigma_1$, and $\sigma_2$, respectively.
			Solid black and dashed purple curves show the total cross section for $M=55$ and $M=2$. Shading indicates the error in the total cross section due to truncation.
		}
		\label{fig:figure_6_resonator_with_shifting}
	\end{figure}
	
	Figure~\ref{fig:figure_6_resonator_with_shifting}(b) shows how the monopole, dipole, and quadrupole cross sections vary with the origin offset $d/R$. These partial cross sections are calculated from Eq.~\eqref{eq:partial_scattering_cross_section} with $M=55$. The total cross section is shown for two truncation orders, $M=2$ and $M=55$.  At $kR=0.42$, the total scattering cross section is independent of the origin offset $d/R$ for both $M=2$ and $M=55$ over the range shown. As the origin moves towards and beyond the slit, the monopole contribution increases and the dipole contribution decreases. At $d/R=2$, the radiation is effectively monopolar. At $kR=1.98$ and $3.24$, the total cross section remains independent of $d/R$ for $M=55$ but varies with the offset for $M=2$. The shaded regions in Fig.~\ref{fig:figure_6_resonator_with_shifting}(b) show how much the $M=2$ calculation underestimates the total cross section relative to the $M=55$ reference. At these higher frequencies, more angular orders are required to recover the same total scattered power at different expansion origins. Therefore, shifting the expansion origin redistributes the scattered power among multipole channels. The total scattering cross section remains unchanged to numerical accuracy when the $T$-matrix retains sufficiently many angular orders. Thus, a suitable choice of the origin can reduce the number of orders needed to achieve a prescribed accuracy~\cite{ustimenko2025optimal}.

	\section{Conclusion}
	
	We have derived a compact analytic $T$-matrix for two-dimensional circular Helmholtz resonators with thin rigid walls and multiple apertures of arbitrary angular widths and positions. The aperture projection operator separates the closed-cylinder response from aperture-induced scattering, and the same aperture equation determines the complex resonance poles. An equivalent formulation combines an aperture-to-Fourier map with an edge-adapted basis for the aperture flux. It converges substantially faster than a direct Fourier expansion and results in scattering spectra in close agreement with full-wave finite-element simulations.
	
	By following the complex poles under variation of the aperture width, we relate changes in resonance frequency and radiative linewidth to features in the scattering spectra and recover the classical narrow-aperture scaling of the fundamental Helmholtz resonance. Comparing single- and double-aperture resonators shows how inversion symmetry controls coupling between multipoles of opposite parity. The inversion-odd part of the $T$-matrix identifies the monopole-dipole channels associated with Willis coupling and their higher-order analogues, such as dipole-quadrupole conversion. When inversion symmetry is broken, internal losses allow scattering and absorption to differ for opposite incidence, while reciprocity ensures equal extinction. Finally, shifting the expansion origin redistributes the scattered power among multipole channels. For a given frequency and illumination, a suitable origin can reduce the number of channels needed to reproduce the total scattering cross section.

	\begin{acknowledgments}
		The authors thank Nikita Ustimenko for fruitful discussions and recommendations. The authors acknowledge financial support from the Russian Science Foundation (25-79-31027).
	\end{acknowledgments}

	\bibliography{References}

\end{document}